## Antenna mechanism of length control of actin cables

Lishibanya Mohapatra[1], Bruce L. Goode[2], Jane Kondev[1]

[1]Department of Physics, Brandeis University, Waltham, MA, USA
[2]Department of Biology and Rosenstiel Basic Medical Sciences Research Center, Brandeis University, Waltham, MA, USA

Actin cables are linear cytoskeletal structures that serve as tracks for myosin-based intracellular transport of vesicles and organelles in both yeast and mammalian cells. In a yeast cell undergoing budding, cables are in constant dynamic turnover yet some cables grow from the bud neck toward the back of the mother cell until their length roughly equals the diameter of the mother cell. This raises the question: how is the length of these cables controlled? Here we describe a novel molecular mechanism for cable length control inspired by recent experimental observations in cells. This "antenna mechanism" involves three key proteins: formins, which polymerize actin, Smy1 proteins, which bind formins and inhibit actin polymerization, and myosin motors, which deliver Smy1 to formins, leading to a length-dependent actin polymerization rate. We compute the probability distribution of cable lengths as a function of several experimentally tuneable parameters such as the formin-binding affinity of Smy1 and the concentration of myosin motors delivering Smy1. These results provide testable predictions of the antenna mechanism of actin-cable length control.

## Introduction

Eukaryotic cells have a complex cytoskeleton that includes vast arrays of microtubules and actin filaments, which governs the internal positioning and movement of cellular substructures such as vesicles and organelles, and dynamic changes in cell polarity, shape, and movement. Many of these processes require the length of the cytoskeletal structures to be tightly controlled. For example, during cell division, the microtubule-based mitotic spindle maintains a remarkably constant size despite undergoing highly dynamic turnover [1–4]. Another example of cellular structures whose lengths are regulated are cilia, which are used for motility and sensation [5–8]. These microtubule-based structures maintain a precise length even though their tubulin building blocks are constantly turning over. Recent studies have begun to address how the length of these microtubule-based structures is maintained [5,7–12]. However, there has been far less attention paid to how the size and length of actin-based structures is determined. The key question that we address here is the mechanism by which the length of actin cables in budding yeast (*Saccharomyces cerevisiae*) is controlled.

Actin is one of the major elements of the cytoskeleton in all eukaryotic cells. It is a protein that polymerizes to form helical two-stranded filaments. The actin cables found in budding yeast cells are estimated to consist of 2-4 filaments bundled in parallel by actin crosslinking proteins. These structures are polymerized by formins[13–16], and serve as tracks for the rapid, directed transport of organelles and vesicles through the mother cell and toward the bud tip. Observations in yeast have shown that during budding, one set of cables is polymerized at the bud neck by the formin Bnr1, which is anchored to a physical scaffold at the bud neck[17]. Bnr1-polymerized actin cables grow into the mother cell, extending toward the rear of the cell, and line the cell cortex [18,19]. As rapidly as the cables grow from the bud neck, they are dismantled at the other end; cables rarely grow past the back of the mother cell, suggesting that their length is regulated [20]. In this paper, we explore theoretically a mechanism of cable length control that acts on the polymerization machinery, formins, which is supported by recent molecular and cellular observations.

Actin cables polymerized by Bnr1 in a yeast cell grow rapidly (~1 μm/s, or ~370 actin subunits/s). Like other formins, Bnr1 remains tightly associated with the fast-growing end of the actin filament [14,21], and thus physically tethers the growing end of the cable to the bud neck while the other end of the cable is disassembled in the cytosol by other cellular factors [22]. The balance of these two antagonistic processes (assembly and disassembly) leads to a steady state cable length. Still, in order to obtain a peaked distribution of cable lengths at steady state, one or both of the rates for assembly and disassembly have to be length dependent. In particular, if the two rates are length independent, and the rate of disassembly (*d*) is greater than the rate of assembly (*r*), then the steady state is characterized by an exponential distribution of lengths. This distribution has a characteristic length given by $\frac{1}{log\left(\frac{d}{r}\right)}$, which is typically small, unless the two rates are almost identical. Therefore, in the absence of a mechanism that leads to a fine balancing of the two rates, the characteristic length is expected to be only a few monomers.

Mechanisms for length dependent depolymerisation have been proposed for microtubule- and actin-based structures. Kinesin motors such as Kip3 and KIF19A move along microtubules and when they reach the end of the track promote dissociation of tubulin subunits, leading to a length-dependent depolymerisation rate [6,9,10,23–25]. In the case of actin, cofilin severs filaments thereby reducing their length in a length-dependent manner. Recently theoretical and experimental studies have shown that this activity alone leads to a peaked distribution of filament lengths in steady state [26–29]. Here we consider an alternative mechanism, in which actin filament length is controlled by negative feedback, which is provided by myosin-motor transport, leading to a length-dependent polymerization rate.

Type-V myosin motors move on cables towards the bud neck and then the bud tip at ~ 3 μm/s, transporting vesicles and other essential cargo destined for the growing bud [19,30]. Recent experiments have shown that Smy1 is a passenger protein of the myosin motor, and is transported to the bud neck where it pauses briefly and is thought to interact with Bnr1, which is anchored there [20]. Further experiments have shown that Smy1 directly binds to Bnr1 and inhibits its actin polymerization activity. As such, when the *SMY1* gene is deleted from cells, a number of the cables grow abnormally long [4]. Here we propose that the active transport of Smy1 along a cable sets up a negative feedback cue to the formin, making the effective cable growth rate length dependent. The length dependence derives from the fact that the rate at which this negative cue is delivered to the formins is proportional to the number of myosin motors bound to and walking on a cable, which serves as an antenna for myosin binding. The goal of this paper is to mathematically explore this antenna mechanism of actin-cable length regulation, and to propose experimental tests of the basic tenets of this model. In particular, we make quantitative predictions for how modulating the strength of the Smy1-formin interaction and the concentration of Smy1 in cells affect the cable-length distribution.

# Results

## Antenna model for cable length control produces a length dependent polymerization rate

The antenna model of actin cable length regulation is based on the idea that a motor delivering an inhibitory cue for polymerization leads to a length dependent growth rate. Smy1 molecules are rapidly transported by myosinV along cables to the barbed ends of the actin filaments in a cable, where they transiently bind to and inhibit the formin (Bnr1). The cable thus acts as a landing pad for myosin+Smy1 inhibitory complexes. Long cables on average encounter more myosin+Smy1 complexes and thereby deliver inhibitory cues at a higher frequency to the formins. This sets up a length dependent negative feedback loop regulating cable elongation rates, and ultimately narrows the distribution of cable lengths in the cell. This antenna model for actin filament length control is related conceptually to the antenna model for a recently-described microtubule length control mechanism, but with a key difference being that in the latter model kinesin motors themselves move directionally on the antenna and upon reaching its end modulate the rate of microtubule disassembly [23,24], whereas in our model the motors carry inhibitors, which upon reaching the end modulate the rate of the actin polymerization engine.

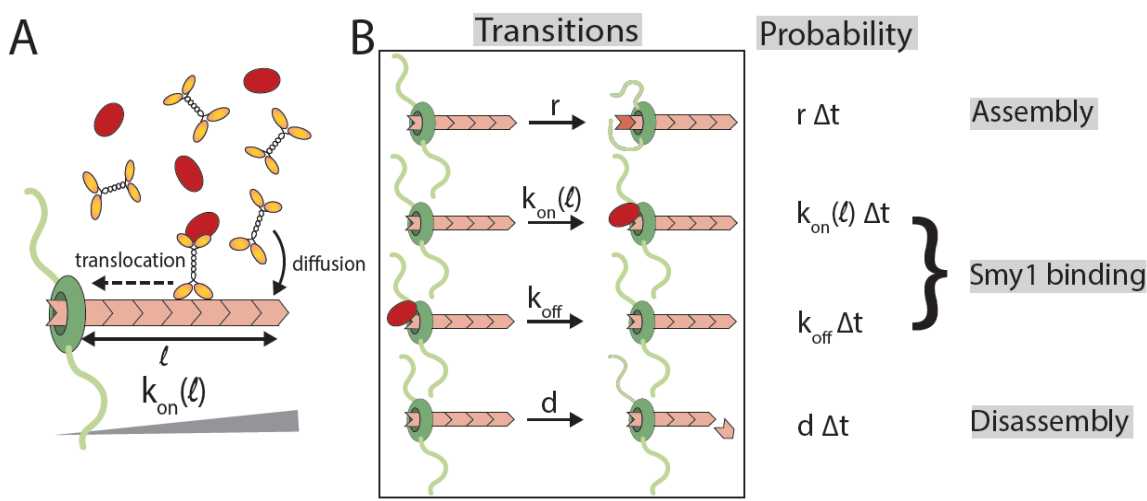

**Figure 1: The antenna model of actin-cable length control**. (A) Smy1 proteins (red) are delivered to the formin (green) at the barbed end of the actin cable by myosin motors (yellow). Smy1 inhibits the polymerization activity of formins upon binding. The directed transport of Smy1 by myosin motors towards the formins leads to a length dependent average assembly rate $k_{on}(l) = wl$; the longer the cables the larger the number of Smy1 proteins delivered the smaller the average assembly rate. (B) A schematic showing all possible transitions between different chemical states in the antenna model. An uninhibited formin assembles cables at a constant rate $r$. Smy1+myosin complexes bind to formin at a rate $k_{on}(l) = wl$ where, $l$ is the length of the cable. Smy1's detach from the formin with a rate $k_{off}$. Regardless of the state of the formin, i.e. whether it has Smy1 bound or not, the filament is disassembled by removal of subunits at a rate $d$.

Here we model the actin cable as a single polymer which grows by the addition of subunits at the formin bound end, and shrinks by subunit removal at the opposite end (Fig 1). Since cables polymerized by Bnr1 in yeast are thought to be comprised of multiple parallel actin filaments bundled together, our model should be taken as an effective description of the assembly and disassembly of this composite structure. In our single-filament model the cable does not grow when Smy1 is inhibiting the formin; subunits are added by the formin at a rate $r$ when the formin is free of Smy1. $k_{off}$ is the rate at which Smy1 molecules detach from the formin, thereby allowing the formin to return to the free/uninhibited state. The rate at which the formin switches from the uninhibited state to the Smy1-bound/inhibited state ($k_{on}$) is equal to the rate of arrival of Smy1 particles to the formin. At steady state, this rate is equal to the rate at which Smy1 proteins diffusing in the cytoplasm are captured by the myosin-carried vesicles (Fig 1A); this assumes that there are no traffic-jams encountered by the myosin motors, which is consistent with our cell experiments and discussed in more detail in the Methods section. According to Smoluchowski, the rate of Smy1 capture is proportional to the Smy1 concentration, and most importantly for our model, the length of the cable, i.e., $k_{on}(l) = wl$. This myosin-dependent delivery of the formin inhibitor Smy1 leads to a length dependent average rate of assembly, which together with a constant disassembly of the cable, which we take to occur by the removal of subunits from the end of the cable at rate $d$, produces a peaked steady-state distribution of cable lengths.

The average time the cable spends in the *on* state, when the formin is active and the cable is growing at rate $r$, is $1/k_{on}(l)$, while the average time the cable spends in the *off* state is $1/k_{off}$. Since we assume that the rate of growth in the *off* state is zero (note that all our conclusions are independent of this assumption as long as the rate of polymerization when Smy1 is bound to formin is smaller than when the formin is free of Smy1), the average rate of polymerization is

$$\bar{r}(l) = r\left(\frac{k_{off}}{k_{off} + k_{on}(l)}\right), \qquad (1)$$

where the factor appearing in parenthesis is the fraction of time that the cable spends in the *on* state. From this calculation we conclude that the average rate of polymerization is length dependent and decreases as the length of a cable increases, since $k_{on}(l) = wl$. Furthermore, the average rate of polymerization depends on the concentration of Smy1 (i.e., $w$ is proportional to [Smy1]) and its binding affinity to the formin ($k_{off}$ is proportional to the dissociation constant), both of which are parameters that can be tuned in experiments.

From the expression for the average rate of polymerization we can compute the steady-state average cable length by equating it with the disassembly rate $d$:

$$\langle l \rangle = \frac{k_{off}}{w}\left(\frac{r}{d} - 1\right). \qquad (2)$$

The key prediction of this equation is that increasing the Smy1 concentration (i.e., increase in $w$) reduces the average cable length, whereas weakening the formin-binding affinity of Smy1 (i.e., increase in $k_{off}$) increases the average cable length. We explore these predictions more thoroughly in the next section. We estimate all four parameters ($r, d, w, k_{off}$) appearing in Equation 2 from in vivo experiments on wild-type yeast cells (see Methods) and study the changes to the cable-length distribution by varying the Smy1 concentration ($w$) and its affinity to formins ($k_{off}$).

## Cable length distribution is regulated by the concentration of Smy1 and its binding affinity to formin

In order to describe the dynamics of an individual cable we mathematically model the antenna mechanism using the master equation formalism. The key quantity to compute is the probability, $P(l,t)$, that the cable has length $l$ (measured here in units of actin subunits) at time $t$. The master equation describes the evolution of $P(l,t)$ in time, by taking into account all the possible changes of the cable state that can occur in a small time interval $\Delta t$ (Fig 1B). For a given cable length, we distinguish between two states depending on whether the formin at its end is inhibited by Smy1 (the *off* state) or free (the *on* state). Therefore we can write $P(l,t) = P_{off}(l,t) + P_{on}(l,t)$, where the probabilities for cable length in the *off* and *on* states satisfy the following master equations (for $l > 0$ and $w, k_{off}, d$ non-zero)

$$\frac{dP_{on}(l,t)}{dt} = r\, P_{on}(l-1) - rP_{on}(l) + d\, P_{on}(l+1) - d\, P_{on}(l) + k_{off}P_{off}(l) - wl\, P_{on}(l)$$

$$\frac{dP_{off}(l,t)}{dt} = d\, P_{off}(l+1) - d\, P_{off}(l) - k_{off}P_{off}(l) + wl\, P_{on}(l). \qquad (3)$$

We use these equations to compute the steady-state distribution of cable lengths $P(l) = P_{on}(l) + P_{off}(l)$, where $P_{on}(l)$ and $P_{off}(l)$ are solutions to Equation 3 when the left-hand sides of these equations are set to zero. The variation of the length distribution with the parameters of the model then provides a stringent set of predictions of the antenna model that can be tested experimentally.

The steady state distribution of cable lengths can be computed exactly using the method of detailed balance in the fast switching regime, i.e., when the rates for switching between the *on* and the *off* states ($k_{on}(l)$ and $k_{off}$) are much greater than the rates of assembly/disassembly. In this limit, the cable can be assumed to have a polymerization rate that is length dependent (see Equation 1) and a disassembly rate $d$. Using the detailed balance condition $P(l)\bar{r}(l) = P(l+1)d$, we obtain

$$P(l) = \left(\frac{r}{d}\right)^l \frac{\left(k_{off}/w\right)^{l-1}}{\left(\frac{\Gamma\left(\frac{k_{off}}{w}+l\right)}{\Gamma(l-1)}\right)} \left(\frac{e^{\frac{k_{off}\,r}{d\,w}} k_{off}\, r(k_{off}-w)\left(\frac{k_{off}\,r}{d\,w}\right)^{-\left(\frac{k_{off}}{w}\right)}\left(\Gamma\left[\frac{k_{off}-w}{w}\right]-\Gamma\left[-1+\frac{k_{off}}{w},\frac{k_{off}\,r}{d\,w}\right]\right)}{d\,w^2}\right)^{-1}$$

(4)

where $\Gamma(x)$ is the Gamma function.

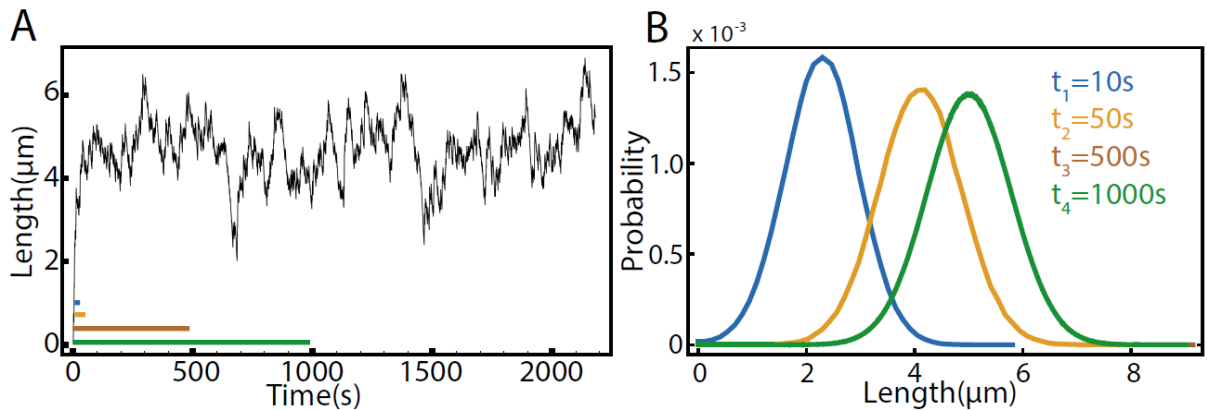

**Figure 2: Time evolution of the cable length distribution.** (A) Time trace of a cable length obtained from simulating the antenna model (see Methods). The parameters used in the simulations are r= 370, $d = 45, w = .004, k_{off} = 1$, all in units of s$^{-1}$. (B) Distribution of filament lengths obtained at different times 10, 50, 500 and 1000 s after the start of the simulation. (Initially the filament length is zero.) We observe that the distribution of lengths settles into the steady state on a time scale of a few hundred seconds. (Note that the brown and green curves corresponding to 500 and 1000 seconds coincide.)

When the rates of switching are comparable to the rates of assembly and disassembly, as is the case for actin cables in budding yeast cells, we are no longer able to obtain an analytic form of the steady state distribution and we resort to numerical simulations of the master equation, Equation 3. We start with a cable of zero length growing from the formin, which acts as a nucleation site. We use the Gillespie algorithm (see Methods) [31,32] to follow the stochastic trajectory in time of the cable length as it polymerizes and depolymerises, while also switching between the *off* and *on* states depending on whether Smy1 is bound to the formin or not. After some time we observe the cable reaching a steady state, when the length distributions no longer changes with time; see Fig 2. For parameter values corresponding to the fast switching regime we find excellent agreement between the stochastic simulations and Equation 4 (see Supplemental Fig S1)[1].

In the slow switching regime, which describes the dynamics of yeast actin cables (see Methods for parameter estimates), we rely solely on the stochastic simulations to obtain steady state distributions of cable lengths for different values of the model parameters.

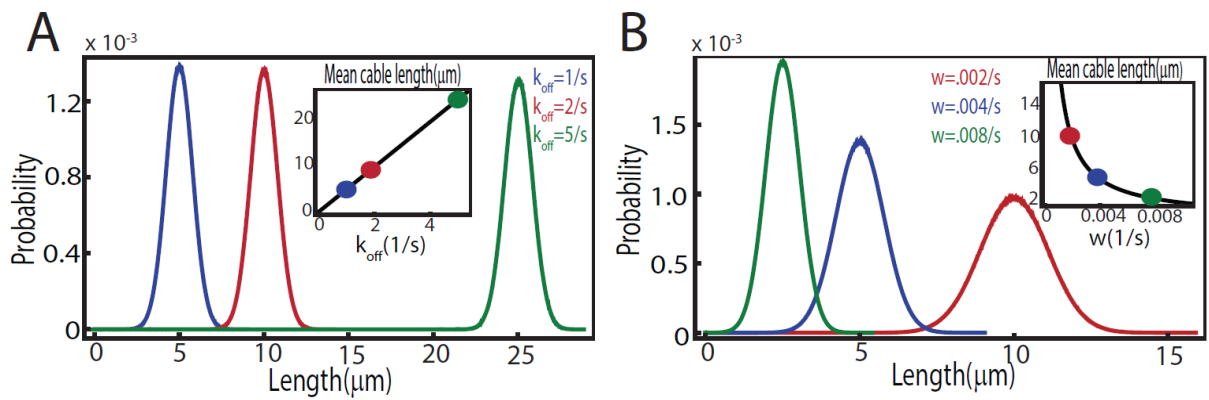

**Figure 3: Steady state cable length distributions depend on Smy1 concentration and binding affinity to formins.** (A) With decreasing Smy1 binding affinity (parametrized by the off rate $k_{off}$) to formins the mean length increases. The inset compares simulation results for the mean cable length and the analytic formula (Equation 2), in black line. (B) With increasing Smy1 concentration (parametrized by the rate $w$) the average length of the cable decreases. Inset shows comparison of simulation results with analytic theory (Equation 2), in black line. The parameter values used in both plots for the polymerization and depolymerization rate are: r=370 s⁻¹ and d=45 s⁻¹ (see Methods). Also in (A) we set $w = 0.004$ s⁻¹ while in (B) we used $k_{off} = 1$ s⁻¹, which are values estimated for these two parameters based on in vivo experiments. In both plots, the blue curves are for estimated parameters for yeast cells.

In Fig 3 we explore the effect of the rate parameters $k_{off}$ and $w$ on the steady state distribution of cable lengths. As explained earlier, the first is proportional to the dissociation constant that measures the binding affinity of Smy1 to formins, while the second rate is proportional to the Smy1 concentration (see Methods for parameter estimates). The results of our simulations for the dependence of the mean cable length on these two parameters are in excellent agreement with Equation 1. Also, in the parameter range explored we observe a difference in the dependence of the width of the steady state length distribution on $k_{off}$ and $w$. Changing the binding affinity of Smy1 to formins has little effect on the width of the length distribution while the Smy1 concentration has a large effect.

---

[1] See Supplementary Information is attached to the main text after the Reference section

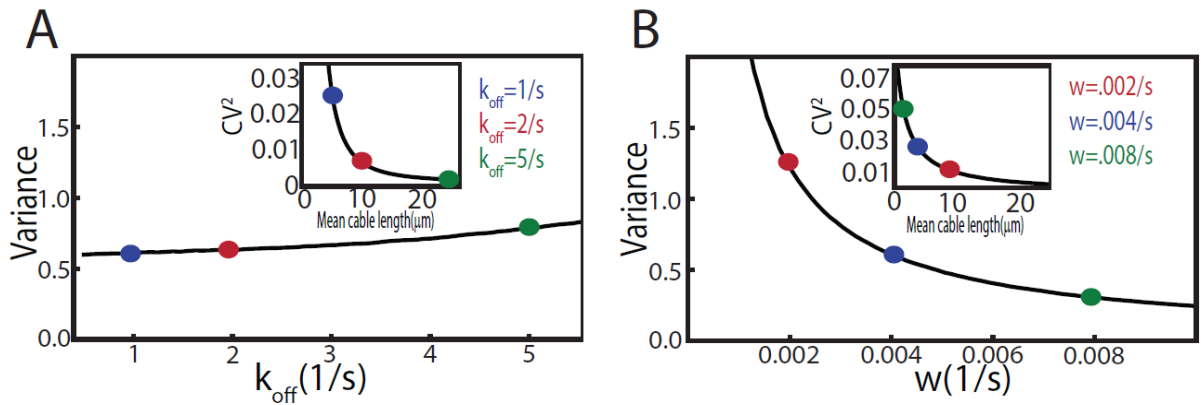

**Figure 4 : Variance in the cable length distribution.** (A) With decreasing Smy1 binding affinity (parametrized by the off rate $k_{off}$) the variance of the cable length distribution slightly increases. The black line was obtained by computing the variance of the cable length distribution by Gillespie simulations for $w$= 0.004 s$^{-1}$ and $k_{off}$ values $0.5 - 5.5$ s$^{-1}$ with a spacing of 0.0625 s$^{-1}$. The inset shows that the noise, as measured by the square of coefficient of variation, decreases. (B) With increasing concentration of Smy1 (parametrized by the rate $w$) the variance increases while the noise (as shown in the inset) decreases. The black line was obtained from simulations using $k_{off} = 1$ s$^{-1}$ and $w = 0.0008 - 0.01$ s$^{-1}$ with a spacing of 0.000125 s$^{-1}$. The parameters $r = 370$ s$^{-1}$ and $d = 45$ s$^{-1}$ are our best estimates for yeast cells (see Methods) also used in Figure 3.

In Fig 4 we show in more detail how the variance and the square of the coefficient of variation ($CV^2 = \frac{variance}{mean^2}$) change as a function of $k_{off}$ and $w$. We see that the square of the coefficient of variation, a standard measure of noise described by a probability distribution, in both cases decreases with increasing average cable length. Fig 3 and Fig 4 also provide a quantitative assessment of how sensitive the length distributions are with respect to the model parameters, in particular the two parameters related to the Smy1 concentration ($w$) and its affinity to formins ($k_{off}$). All the plots shown in Fig 3 and Fig 4 constitute specific predictions of the antenna model, which can be readily tested by in vitro experiments. While more difficult, experiments in vivo in which these two parameters are varied and the change of cable length distribution is measured, are also possible.

## The mean and variance of the cable length distribution can be controlled independently

The key feature of the antenna model proposed here is the switching of the cable between two states, one in which the formin is active and the cable is growing, and the other in which the formin is inactive (by virtue of Smy1 being bound to it) and the cable is therefore shrinking. The balance of the two states leads to the average cable length given in Equation 2. The same average length can be achieved either by large $k_{off}$ and $w$, or by small $k_{off}$ and $w$, as long as the average rate of polymerization, Equation 1, is the same. In other words, the same mean length can be achieved either by having a small concentration of Smy1 proteins present in solution but they stay bound to the formin for a longer time, or in the alternate case where a large number of Smy1 proteins are in solution, but they associate with formin for a shorter time.

The width of the length distribution, on the other hand, will not be the same in these two extremes. When the switching rates are slow, we expect that the formins will spend long periods of time in the active and the inactive state leading to large fluctuations in the cable length, when compared to the situation when the switching is fast. This leads to the possibility that by tuning the concentration of Smy1 and its binding affinity to formins one is able to control the mean and the width of the length distribution independently. These properties distinguish the antenna mechanism discussed here from most other models of length control described previously. Interestingly, and roughly related to our findings, different versions of the antenna model of microtubule length control, which lead to the same mean microtubule length, have been reported to predict dramatically different steady-state fluctuations [34].

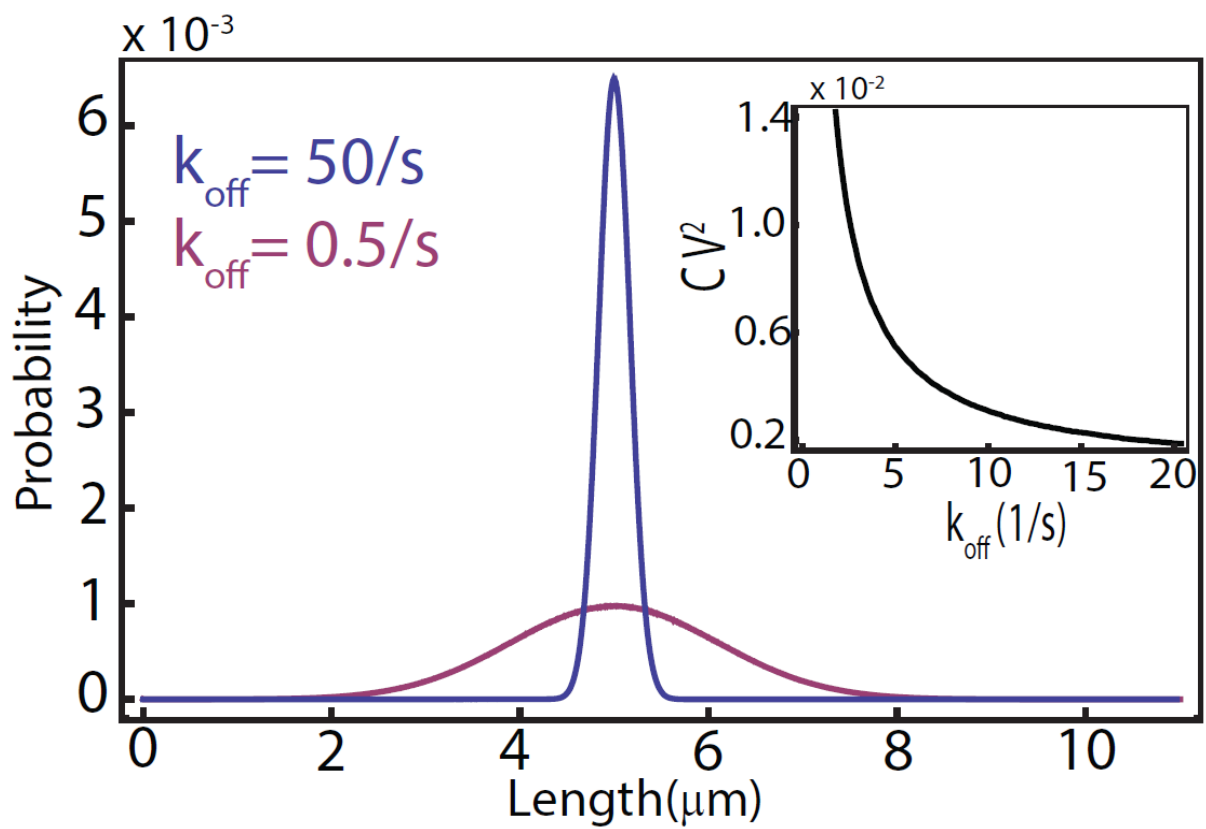

**Figure 5: The mean and variance of the cable length can be independently controlled within the antenna model.** The same mean cable length (5 microns), is obtained either by a combination of a large Smy1-formin binding affinity and a small Smy1 concentration (parametrized by $k_{off}$ and $w$ respectively), or by a weak affinity and large concentration. The distribution in the weak affinity case (blue) is sharper than in the strong affinity case (red). Parameters used for the polymerization and depolymerisation rate were $r = 370$ s$^{-1}$, $d = 45$ s$^{-1}$, $k_{off} = 0.5$ s$^{-1}$(red) and 50 s$^{-1}$(blue) ; for chosen values of $k_{off}$, the rate $w$ was calculated from Equation 2 for the mean length. Inset: Square of the coefficient of variation of the cable length distribution (measured by variance/mean$^2$) decreases with $k_{off}$ when the mean length is kept fixed by adjusting $w$. Parameters used for the simulations (in black line) were $r = 370$ s$^{-1}$, $d = 45$ s$^{-1}$, $w = 0.004$ s$^{-1}$ and $k_{off}$ was varied from 0.5-20 s$^{-1}$ in steps of 0.125 s$^{-1}$.

In order to test our expectations about how the variance and the mean of the cable length distribution can be controlled separately, we computed the distributions for different values of the rates $k_{off}$ and $w$ while keeping their ratio the same; in accordance with Equation 2 this guarantees that the mean length is fixed. We also kept the rate of assembly $r$ and the rate of disassembly $d$ fixed as we do not expect these to change when tuning the concentration of Smy1 and its binding affinity to the formin. Using a Gillespie simulation of the master equation (Equation 3) we obtained length distributions for the slow and fast switching cases; see Fig 5. As expected, we observe more noise (larger width for the same mean) in the slow switching situation, which could be realized experimentally by having a small concentration of Smy1 mutants with a large binding affinity for formins. The decrease in the square of coefficient of variation of the length distribution with decreasing binding affinity of Smy1 is shown in the inset to Fig 5.

## Discussion

Actin-binding proteins play a multitude of critical roles in maintaining the shape and dynamics of different actin structures, including the polarized actin cables found in yeast cells that support intracellular transport and asymmetric cell division. In this paper, we describe an antenna mechanism by which formins, along with the myosin-delivered formin inhibitor Smy1, control the cable length by making the rate of cable polymerization length dependent. Our model predicts that shorter cables grow faster than long cables, as they are subject to less inhibition by Smy1. We compute length distributions as a function of model parameters that can be tuned experimentally by changing the Smy1 concentration and its binding affinity to the formin. Interestingly, we observed that the mean cable length and variance can be independently controlled within the antenna mechanism by tuning these two model parameters simultaneously. Our results provide quantitative predictions for future in vivo experiments aimed at testing the correlation between cable length and growth rate, and in vitro experiments aimed at reconstituting cable assembly with length control feedback from purified proteins (actin, formin, Smy1 and myosinV).

The antenna model described here assumes a constant supply of free actin monomers in solution. This is a reasonable assumption for an in vitro experiment where the amount of actin monomers taken up by cables is small, but it might not hold for the in vivo situation. In vivo, even in the absence of Smy1 we expect cable growth to be dependent on processes that contribute to actin monomer recycling, and thus on factors that affect the disassembly rate, $d$. Furthermore, in vivo, proteins that sever actin filaments may provide an additional mechanism of length control. Here we make estimates to address the role that the finite monomer pool and severing may play in cable length regulation in wild-type yeast cells.

## Mechanisms of cable length control in vivo

An alternative length control mechanism to the antenna mechanism, discussed above, is the finite supply of actin monomers in a cell [35]. As the cables grow, the free actin concentration decreases, leading to a decrease in the polymerization rate of actin filaments that make up the cables. When the polymerization rate equals the disassembly rate, steady state is reached. However, below we make estimates that suggest that the finite monomer pool cannot be the only source of length regulation in vivo, and this is supported by the observation that some of the cables overgrow in cells when *SMY1* is deleted [20].

In the presence of a finite monomer pool, the average polymerization rate can be estimated as $r'\,(N - N_c\,D\,\langle n \rangle)$, where $N$ is the total number of actin molecules in the cell (in both filamentous and monomeric forms), $N_c$ is the number of cables and $r'$ is the assembly rate of free monomer; note than in the absence of cables, when all of the actin molecules are in monomeric form, $r' = r/N$ . Here, for the purposes of an estimate, we assume a simple geometry for the cables, where each cable has an average length $\langle n \rangle$, and consists of $D$ actin filaments in parallel bundled together. In steady state, the average polymerization rate is equal to the depolymerisation rate $d$, which leads to an average cable length $\langle n \rangle = (N - d/r')/N_c D$.

The total number of actin molecules in the mother-cell (which contains the cables of interest) can be estimated by considering the concentration of actin in the cell's cytoplasm, which we have measured by quantitative western blotting, and multiplying it by the known volume of a yeast mother-cell, $N = 10 \; \mu M \times \frac{4\pi}{3}(2.5 \; \mu m)^3 = 3 \times 10^5$ actin proteins. Observations in vivo suggest that the number of cables is roughly 10 and they have a thickness of about $D$=4 filaments. Furthermore, if we take into account the in vivo rates of cable assembly and disassembly, $r = 370 \frac{1}{s}$, $r' = 1.2 \times 10^{-3} \frac{1}{s}$, $d = 45 \frac{1}{s}$, we estimate an average cable length of $\langle n \rangle = 18 \; \mu m$ (using the conversion 1 $\mu m$ = 370 monomers). (This estimate doesn't take into account actin patches as there are relatively few of these structures in the mother cell.) The estimated average cable length is more than a factor of three longer than what is observed in wild-type yeast cells, suggesting the presence of additional length-control mechanisms. Interestingly enough in mutant cells lacking Smy1, we observe some cables whose length is roughly twice that seen in wild type cells; this observation is consistent with the idea that the finite monomer pool limits cable length in the absence of the Smy1-dependent antenna mechanism.

Another process that can control cable length is actin severing, in which proteins like cofilin bind to the sides of filaments and induce breaks. This leads to the breaking off of polymer fragments, which are rapidly capped and depolymerized since they no longer have formins at their ends [27,29]. Since filaments within a cable provide binding sites for cofilin, the longer the cable, the higher the rate of cofilin binding. This may lead to a length-dependent severing rate, $sl$, where $s$ is the severing rate per micron of cable per second. Since cables are anchored at the bud-neck, when a cable gets severed (by the severing of constitutive filaments), approximately and on average half of the subunits are lost, i.e., they are no longer part of the cable attached to the bud-neck. Therefore, assuming that severing can occur at any position along the cable that cofilin binds to, the depolymerisation rate (i.e., rate of subunit loss) becomes length dependent, $d(l) = sl \times \frac{l}{2} = \frac{sl^2}{2}$. To obtain the steady state filament length we set this depolymerisation rate equal to the polymerization rate, which leads to the formula $\langle l \rangle = \sqrt{2r/s}$. Taking our estimated value for the polymerization rate, $r = 1 \mu m/s$, and the maximum in vitro measured severing rate (at 10 nM cofilin) $s = 10^{-3} \mu m^{-1} \; s^{-1}$ [36], the estimate of the steady state cable length is $\langle l \rangle = 45 \mu m$, more than five times the length observed in vivo. We expect this estimate to be in fact a lower bound on the average length obtained by the severing mechanism, since the optimized severing rate used above is actually decreased at both lower and higher concentrations of cofilin [36]. Therefore this estimate suggests that severing cannot be the only mechanism of length control.

It should be noted that in our consideration of the effects of severing on cable length control we only consider severing by cofilin. However, in cells there are a number of other co-factors that work with cofilin (e.g., coronin, Srv2/CAP, Aip1) and are likely to increase the rate of severing to further reduce cable length [37–39]. Hard numbers for the contributions of these co-factors to severing are not yet available, but once they are, they can be worked into this model. Another key factor is the presence of Tropomyosin proteins coating the cables. Tropomyosin is essential for cable formation [40,41], and is thought to protect cables at least temporarily from cofilin-mediated severing. Thus, Tropomyosin may direct cofilin-mediated severing to the 'older' ends of the cables, which is consistent with the model of dissociation that we have adopted for the antenna mechanism.

The above estimates suggest that cable lengths in vivo cannot be controlled by the finite actin monomer pool and severing alone, and requires additional length-dependent feedback mechanisms. This is consistent with our cell experiments in which we observe striking changes in cable lengths upon deletion of *SMY1*[20]. This raises the intriguing possibility that cells have evolved multiple mechanisms of cable-length control, including several other potential ones besides Smy1. For example, the specific conformation that F-actin adopts in different nucleotide states is likely to affect severing along cables, and therefore any protein that decorates cables and alters the nucleotide state and/or conformation of F-actin could be part of an additional length control mechanism [42]. In addition, the ends of overgrown cables colliding with the cell cortex might change the mechanical stress of a cable leading to a change in its assembly or disassembly rate. Further, the mechanical strain on filaments induced by myosin action can affect severing by cofilin [43] and therefore alter the disassembly rate.

In this paper we focused on cables assembled by only one of the two budding yeast formins, Bnr1, which is stably anchored to the bud neck [17]. However, the other budding yeast formin, Bni1, is highly distinct in its cellular dynamics. Bni1 molecules appear to be transiently recruited to the bud tip to assemble cables, then released, similar to the formin For3 in fission yeast[17,44,45] . A recent study of For3 discussed how this formin might control cable length in fission yeast [44,45]. Their model considered the transient association of For3 with the cell tip leading to the assembly of actin filaments by the formin. For3 and the newly polymerized actin filaments are then released from the cell tip and carried passively into the cell interior by the retrograde flow of actin filaments in the cable. Upon release from the cell cortex, the actin filaments in cables can disassemble, increasing the amount of free actin which, in turn, increases For3 dissociation from the cell tip. This coupling between actin monomer levels and For3 attachment leads to a steady state at realistic values of rate constants and actin and For3p concentrations [44].Whether or not a similar length control mechanism is employed for Bni1 generated cables in budding yeast is an intriguing open question.

## Effect of myosin motor speed and processivity on cable length control

In our model we assume for simplicity that myosin motors transporting Smy1 to the anchored formins do not fall off the cables. Furthermore, it is assumed that the rate of delivery of Smy1 by myosin is greater than the polymerization rate of the cable. Both conditions are necessary for every Smy1 molecule captured by the actin-cable 'antenna' to be delivered to the formins. Here we address the experimental evidence for these two assumptions.

In wild type cells, Smy1-GFP was directly observed to be trafficked by the myosin motor and delivered, uninterrupted, to the formin [20]. Smy1 is on vesicles, which have multiple myosin motors attached to them, which may explain why processivity does not seem to be an issue in vivo, and validates the assumption in our model that delivery of Smy1 is uninterrupted. Also, in a wild type cell, the observed anterograde transport rate of vesicles toward the bud neck is 3 μm/s [20], which, given a retrograde elongation rate of cables of 0.5-1 μm/s [19], suggests a myosin motor speed of about 3.5- 4 μm/s. These observations are consistent with the assumption that the rate of transport of Smy1 toward the formin is much greater than the rate of cable elongation. Further, this predicts that the antenna mechanism would not be effective for controlling cable length if the myosin speed was less than 1 μm/s since in that case Smy1 will not be delivered to the formin. This qualitative prediction can be tested using myosin mutants [46] with reduced in vivo transport speeds.

Another interesting point to consider is the wide range of cable elongation rates reported in the literature, ranging between a few tenths of a micron per second to several microns per second [18,19]. The antenna model provides a possible explanation for this observation. Namely, the model predicts that the cable extension rate decreases with the cable length (Equation 1). Therefore, it is possible that the range of reported cable elongation rates is due to the variability of the lengths of cables whose extension rate was measured.

In conclusion, the antenna model involving formins, Smy1 and myosin motors, is a novel molecular mechanism for length control of actin cables, which we have proposed based on experimental evidence in living cells. While in cells it is almost certain that multiple mechanisms contribute to cable length control, in vivo observations as well as theoretical estimates indicate that the antenna mechanism is an important factor in controlling the length of these actin structures. Here we have explored this model theoretically, and made a number of predictions that can be tested in vivo, and in vitro using a reconstituted system consisting of purified actin, formin, myosin and Smy1. In particular, we compute the effect of changing the concentration of Smy1 and its binding affinity to formin on the distributions of cable lengths. Therefore quantitative measurements of this distribution in an in vitro reconstituted system of length control would serve as a stringent test of the antenna mechanism. An interesting qualitative prediction of the model is that the variability and the mean of the actin cable length can be tuned independently by simultaneously tuning these two control parameters (Smy1 concentration, and Smy1 affinity to formin). Whether such differential control is something that is used by cells to tune the length of actin cables is an interesting open question.

## Methods

### Estimation of model parameters

The antenna mechanism is specified by four parameters, which can be estimated based on published experiments. In fact, there are two published studies that measured rates of cable growth. In an earlier study, Pon and colleagues measured the rate to be ~ 0.3-0.6 μm/s [18]. In a later study, Wedlich-Soldner and colleagues used improved methods for imaging and quantifying cable growth rates (employing TIRF microscopy in vivo) and reported rates of ~ 1μm/s [19]. The value $r = 1$μm/s (for the polymerization rate when the formin is free of Smy1) we have adopted is based on the observed maximum rate of cable growth in vivo [19]; in making this estimate we assume that the maximum growth rate corresponds to small cables for which the attenuation of growth by Smy1 is not significant and therefore the average polymerization rate is much greater than the depolymerisation rate and is roughly equal to the observed growth rate of the cable. This value for the growth rate has also been independently confirmed by TIRF microscopy in our own lab (Julian Eskin and B.G., unpublished data).

In cell experiments GFP labelled Smy1 proteins are seen to pause at the bud neck for about a second in wild type cells [20] and so we estimate $k_{off} =1$/s for the rate of Smy1 falling off of the formins.

The myosin-aided delivery rate of Smy1 to the formin, leads to a length dependent on rate $k_{on}(l) = wl$. We estimate the value of the parameter $w$ using the observed number of myosin+Smy1 complexes on the cable. If we model the actin cable as a polymer with $l$ subunits, at every subunit we can consider all the processes by which the myosin+Smy1 complexes arrive and depart the particular subunit. In steady state the number of complexes arriving and departing need to balance. In particular, myosin+Smy1 can either reach the $x^{th}$ subunit ($1 < x < l$) diffusively from the cell cytosol with a rate $k_{on}^0$ (which is proportional to the concentration of Smy1 proteins), or by translocating from the $x - 1$ subunit, with a rate $v$. We assume that the motors do not fall off the polymer and therefore the only way that they leave the $x^{th}$ subunit is by translocating to subunit $x + 1$. At steady state, the number of complexes arriving and departing the $x^{th}$ subunit are equal and therefore the steady state number is $N(x) = \frac{x\,k_{on}^0}{v}$ [24]. Using this quantity we can compute the total number of motors (myosin+Smy1 complexes) on the polymer (or cable) by summing over all subunits:

$$N_{tot} = \sum_{x=0}^{l} N(x) = \frac{k_{on}^0}{v} \frac{l(l+1)}{2}. \quad (5)$$

The rate of delivery of Smy1 to the formin at the barbed end is equal to the number of complexes that translocate from the $l^{th}$ subunit to the formin, i.e. $k_{on}(l) = vN(l) = lk_{on}^0$; therefore $k_{on}^0$ is equal to the previously defined parameter $w$. Using Equation 5 we can solve for $k_{on}^0$, to obtain the relation $N_{tot} = w\frac{L(L+L_0)}{2L_0 V}$, where $V = vL_0$, is the myosin velocity in units of microns per second, and $L = l\,L_0$ is the cable length in microns; $L_0 = 2.7$ nm is the size of an actin subunit in the cable. In our cell experiments, we observe that that $N_{tot} = 5$, $L = 5$ μm, and $V = 3.5$ μm/s which yields $w = 0.004$ s$^{-1}$. In making these estimates, we do not consider the possibility of the density of myosin+Smy1 complexes on cables reaching saturation. This is supported by our live-cell imaging of secretory vesicles (marked with GFP fusions to either Sec4 or Smy1), which show that vesicles never experience traffic jams. Instead, single vesicles processively move along cables and reach the bud neck uninterrupted.

We use these the above estimated values for the three parameters ($r, k_{off}, w$) and the expression for mean cable length (Equation 2) to obtain a value of the fourth parameter, the depolymerisation rate $d$. By equating the mean cable length to 5 microns, which is the typical cable length we observe in vivo, and using the parameter values listed above, we estimate the depolymerisation rate $d = 0.12$ µm/s or 45 subunits/s.

It is important to note that while our estimates for the model parameters are quite rough our conclusions about the effect of Smy1 concentration and its affinity to formins on the distribution of cable lengths are independent of the particular parameter values.

## Simulation protocol

In order to solve the master equations in the parameter regime corresponding to actin cable growth in wild type yeast cells, we resorted to numerical simulations. We start with a cable of zero length and then use the Gillespie algorithm [31,32] to follow the stochastic trajectory of a cable. In the simulation the state of the system is characterized by the cable length and whether the formin is active (free of Smy1) or inactive (Smy1 bound). In one step of the simulation we choose one of the set of all possible transitions from the current state of the system to the next. The transitions are chosen at random according to their relative weight, which is proportional to the rate of the transition. Once a particular transition is chosen the system is updated to a new state, which becomes the new current state. The time elapsed between two consecutive transitions is drawn from an exponential distribution, the rate parameter of which equals the sum of all the rates of allowed transitions. This process is repeated for a long enough time such that the length of the cable reaches steady state; see Fig 2A. We obtain many such trajectories of a single cable and then compute the steady state distributions of length and the first and second moments of the distribution for the mean and variance of cable lengths.

## Acknowledgements

We would like to thank the other Brandeis Cable Club members Julian Eskin, Brian Graziano, and Sal Alioto, Nenad Pavin and Matko Gluncic for stimulating discussions.

## References

1. Dumont S, Mitchison TJ (2009) Force and Length in the Mitotic Spindle. Curr Biol 19: R749–R761. doi:10.1016/j.cub.2009.07.028.

2.  Goshima G, Wollman R, Stuurman N, Scholey JM, Vale RD (2005) Length Control of the Metaphase Spindle. Curr Biol 15: 1979–1988. doi:10.1016/j.cub.2005.09.054.

3.  Rizk RS, Discipio KA, Proudfoot KG, Gupta ML Jr (2014) The kinesin-8 Kip3 scales anaphase spindle length by suppression of midzone microtubule polymerization. J Cell Biol 204: 965–975. doi:10.1083/jcb.201312039.

4.  Wang H, Brust-Mascher I, Cheerambathur D, Scholey JM (2010) Coupling between microtubule sliding, plus-end growth and spindle length revealed by kinesin-8 depletion. Cytoskelet Hoboken NJ 67: 715–728. doi:10.1002/cm.20482.

5.  Ishikawa H, Marshall WF (2011) Ciliogenesis: building the cell's antenna. Nat Rev Mol Cell Biol 12: 222–234. doi:10.1038/nrm3085.

6.  Niwa S, Nakajima K, Miki H, Minato Y, Wang D, et al. (2012) KIF19A Is a Microtubule-Depolymerizing Kinesin for Ciliary Length Control. Dev Cell 23: 1167–1175. doi:10.1016/j.devcel.2012.10.016.

7.  Marshall WF, Qin H, Rodrigo Brenni M, Rosenbaum JL (2005) Flagellar length control system: testing a simple model based on intraflagellar transport and turnover. Mol Biol Cell 16: 270–278. doi:10.1091/mbc.E04-07-0586.

8.  Marshall WF, Rosenbaum JL (2001) Intraflagellar transport balances continuous turnover of outer doublet microtubules: implications for flagellar length control. J Cell Biol 155: 405–414. doi:10.1083/jcb.200106141.

9.  Gardner MK, Zanic M, Gell C, Bormuth V, Howard J (2011) Depolymerizing Kinesins Kip3 and MCAK Shape Cellular Microtubule Architecture by Differential Control of Catastrophe. Cell 147: 1092–1103. doi:10.1016/j.cell.2011.10.037.

10. Hough LE, Schwabe A, Glaser MA, McIntosh JR, Betterton MD (2009) Microtubule Depolymerization by the Kinesin-8 Motor Kip3p: A Mathematical Model. Biophys J 96: 3050–3064. doi:10.1016/j.bpj.2009.01.017.

11. Avasthi P, Marshall WF (2012) Stages of Ciliogenesis and Regulation of Ciliary Length. Differ Res Biol Divers 83: S30–S42. doi:10.1016/j.diff.2011.11.015.

12. Engel BD, Ishikawa H, Feldman JL, Wilson CW, Chuang P-T, et al. (2011) A cell-based screen for inhibitors of flagella-driven motility in Chlamydomonas reveals a novel modulator of ciliary length and retrograde actin flow. Cytoskelet Hoboken NJ 68: 188–203. doi:10.1002/cm.20504.

13. Kovar DR, Pollard TD (2004) Insertional assembly of actin filament barbed ends in association with formins produces piconewton forces. Proc Natl Acad Sci U S A 101: 14725–14730. doi:10.1073/pnas.0405902101.

14. Kovar DR, Harris ES, Mahaffy R, Higgs HN, Pollard TD (2006) Control of the Assembly of ATP- and ADP-Actin by Formins and Profilin. Cell 124: 423–435. doi:10.1016/j.cell.2005.11.038.

15. Vavylonis D, Kovar DR, O'Shaughnessy B, Pollard TD (2006) Model of formin-associated actin filament elongation. Mol Cell 21: 455–466. doi:10.1016/j.molcel.2006.01.016.

16. Mizuno H, Higashida C, Yuan Y, Ishizaki T, Narumiya S, et al. (2011) Rotational movement of the formin mDia1 along the double helical strand of an actin filament. Science 331: 80–83. doi:10.1126/science.1197692.

17. Buttery SM, Yoshida S, Pellman D (2007) Yeast formins Bni1 and Bnr1 utilize different modes of cortical interaction during the assembly of actin cables. Mol Biol Cell 18: 1826–1838. doi:10.1091/mbc.E06-09-0820.

18. Yang H-C, Pon LA (2002) Actin cable dynamics in budding yeast. Proc Natl Acad Sci 99: 751–756. doi:10.1073/pnas.022462899.

19. Yu JH, Crevenna AH, Bettenbühl M, Freisinger T, Wedlich-Söldner R (2011) Cortical actin dynamics driven by formins and myosin V. J Cell Sci 124: 1533–1541. doi:10.1242/jcs.079038.

20. Chesarone-Cataldo M, Guérin C, Yu JH, Wedlich-Soldner R, Blanchoin L, et al. (2011) The Myosin Passenger Protein Smy1 Controls Actin Cable Structure and Dynamics by Acting as a Formin Damper. Dev Cell 21: 217–230. doi:10.1016/j.devcel.2011.07.004.

21. Goode BL, Eck MJ (2007) Mechanism and function of formins in the control of actin assembly. Annu Rev Biochem 76: 593–627. doi:10.1146/annurev.biochem.75.103004.142647.

22. Moseley JB, Goode BL (2006) The Yeast Actin Cytoskeleton: from Cellular Function to Biochemical Mechanism. Microbiol Mol Biol Rev 70: 605–645. doi:10.1128/MMBR.00013-06.

23. Varga V, Leduc C, Bormuth V, Diez S, Howard J (2009) Kinesin-8 Motors Act Cooperatively to Mediate Length-Dependent Microtubule Depolymerization. Cell 138: 1174–1183. doi:10.1016/j.cell.2009.07.032.

24. Varga V, Helenius J, Tanaka K, Hyman AA, Tanaka TU, et al. (2006) Yeast kinesin-8 depolymerizes microtubules in a length-dependent manner. Nat Cell Biol 8: 957–962. doi:10.1038/ncb1462.

25. Melbinger A, Reese L, Frey E (2012) Microtubule Length Regulation by Molecular Motors. Phys Rev Lett 108: 258104. doi:10.1103/PhysRevLett.108.258104.

26. Pavlov D, Muhlrad A, Cooper J, Wear M, Reisler E (2007) ACTIN FILAMENT SEVERING BY COFILIN. J Mol Biol 365: 1350–1358. doi:10.1016/j.jmb.2006.10.102.

27. Reymann A-C, Suarez C, Guérin C, Martiel J-L, Staiger CJ, et al. (2011) Turnover of branched actin filament networks by stochastic fragmentation with ADF/cofilin. Mol Biol Cell 22: 2541–2550. doi:10.1091/mbc.E11-01-0052.

28. Suarez C, Roland J, Boujemaa-Paterski R, Kang H, McCullough BR, et al. (2011) Cofilin Modulates the Nucleotide State of Actin Filaments and Severs at Boundaries of

Bare and Decorated Segments. Curr Biol CB 21: 862–868. doi:10.1016/j.cub.2011.03.064.

29. Roland J, Berro J, Michelot A, Blanchoin L, Martiel J-L (2008) Stochastic Severing of Actin Filaments by Actin Depolymerizing Factor/Cofilin Controls the Emergence of a Steady Dynamical Regime. Biophys J 94: 2082–2094. doi:10.1529/biophysj.107.121988.

30. Bretscher A (2003) Polarized growth and organelle segregation in yeast the tracks, motors, and receptors. J Cell Biol 160: 811–816. doi:10.1083/jcb.200301035.

31. Gillespie DT (1976) A general method for numerically simulating the stochastic time evolution of coupled chemical reactions. J Comput Phys 22: 403–434. doi:10.1016/0021-9991(76)90041-3.

32. Gillespie DT (1977) Exact stochastic simulation of coupled chemical reactions. J Phys Chem 81: 2340–2361. doi:10.1021/j100540a008.

33. Miyake A, Stockmayer WH (1965) Theoretical reaction kinetics of reversible living polymerization. Makromol Chem 88: 90–116. doi:10.1002/macp.1965.020880107.

34. Kuan H-S, Betterton MD (2013) Biophysics of filament length regulation by molecular motors. Phys Biol 10: 036004. doi:10.1088/1478-3975/10/3/036004.

35. Goehring NW, Hyman AA (2012) Organelle growth control through limiting pools of cytoplasmic components. Curr Biol CB 22: R330–R339. doi:10.1016/j.cub.2012.03.046.

36. Andrianantoandro E, Pollard TD (2006) Mechanism of Actin Filament Turnover by Severing and Nucleation at Different Concentrations of ADF/Cofilin. Mol Cell 24: 13–23. doi:10.1016/j.molcel.2006.08.006.

37. Gandhi M, Achard V, Blanchoin L, Goode BL (2009) Coronin switches roles in actin disassembly depending on the nucleotide state of actin. Mol Cell 34: 364–374. doi:10.1016/j.molcel.2009.02.029.

38. Kueh HY, Charras GT, Mitchison TJ, Brieher WM (2008) Actin disassembly by cofilin, coronin, and Aip1 occurs in bursts and is inhibited by barbed-end cappers. J Cell Biol 182: 341–353. doi:10.1083/jcb.200801027.

39. Balcer HI, Goodman AL, Rodal AA, Smith E, Kugler J, et al. (2003) Coordinated regulation of actin filament turnover by a high-molecular-weight Srv2/CAP complex, cofilin, profilin, and Aip1. Curr Biol CB 13: 2159–2169.

40. Kuhn TB, Bamburg JR (2008) Tropomyosin and ADF/cofilin as collaborators and competitors. Adv Exp Med Biol 644: 232–249.

41. Skau CT, Kovar DR (2010) Fimbrin and tropomyosin competition regulates endocytosis and cytokinesis kinetics in fission yeast. Curr Biol CB 20: 1415–1422. doi:10.1016/j.cub.2010.06.020.

42. Erlenkämper C, Kruse K (2009) Uncorrelated changes of subunit stability can generate length-dependent disassembly of treadmilling filaments. Phys Biol 6: 046016. doi:10.1088/1478-3975/6/4/046016.

43. Wiggan O, Shaw AE, DeLuca JG, Bamburg JR (2012) ADF/cofilin regulates actomyosin assembly through competitive inhibition of myosin II binding to F-actin. Dev Cell 22: 530–543. doi:10.1016/j.devcel.2011.12.026.

44. Wang H, Vavylonis D (2008) Model of For3p-Mediated Actin Cable Assembly in Fission Yeast. PLoS ONE 3: e4078. doi:10.1371/journal.pone.0004078.

45. Martin SG, Chang F (2006) Dynamics of the formin for3p in actin cable assembly. Curr Biol CB 16: 1161–1170. doi:10.1016/j.cub.2006.04.040.

46. Hodges AR, Bookwalter CS, Krementsova EB, Trybus KM (2009) A non-processive class V myosin drives cargo processively when a kinesin-related protein is a passenger. Curr Biol CB 19: 2121–2125. doi:10.1016/j.cub.2009.10.069.

## Supplementary Information

## Analytical solution of the master equation

We used detailed balance to solve the master equation in the regime where the switching rates (parameters $k_{off}$ and $w$) are much larger than rates of assembly and disassembly (parameters $r$ and $d$). Using the detailed balance condition, $P(l)\bar{r}(l) = P(l+1)d$, where $\bar{r}(l)$ is the average polymerization rate (see Equation 1), we obtain (for $l > 0$ and $w, k_{off}, d$ non-zero)

$$P(l)\, r\, \frac{k_{off}}{k_{off} + wl} = d\, P(l+1)$$

By solving this expression recursively we can express $P_l$ in terms of $P_0$, the probability of zero subunits present at the formin,

$$P(l) = \left(\frac{r}{d}\right)^l \prod_{i=0}^{l-1} \left(\frac{k_{off}}{k_{off} + i\,w}\right) P_0.$$

We use the normalization condition for $P(l)$ to obtain $P_0$, which than gives us an simple analytic formula for the length distribution

$$P(l) = \left(\frac{r}{d}\right)^l \frac{\left(k_{off}/w\right)^{l-1}}{\left(\frac{\Gamma\left(\frac{k_{off}}{w}+l\right)}{\Gamma(l-1)}\right)} \left(\frac{e^{\frac{k_{off}\,r}{d\,w}} k_{off}\, r(k_{off}-w)\left(\frac{k_{off}\,r}{d\,w}\right)^{-\left(\frac{k_{off}}{w}\right)}\left(\Gamma\left[\frac{k_{off}-w}{w}\right]-\Gamma\left[-1+\frac{k_{off}}{w},\frac{k_{off}\,r}{d\,w}\right]\right)}{d\,w^2}\right)^{-1}.$$

where $\Gamma(x)$ is the Gamma function.

We compare this formula to results of simulations in Figure S1. In the fast switching regime where the formula is expected to be valid, we observe agreement with the distribution obtained from simulations. In the regime of slow switching i.e where the switching rates (parameters $k_{off}$ and $w$) are much smaller than rates of assembly and disassembly (parameters $r$ and $d$), the distribution derived from the analytical expression is much narrower than that obtained from simulations, consistent with our intuition that slow switching increases noise.

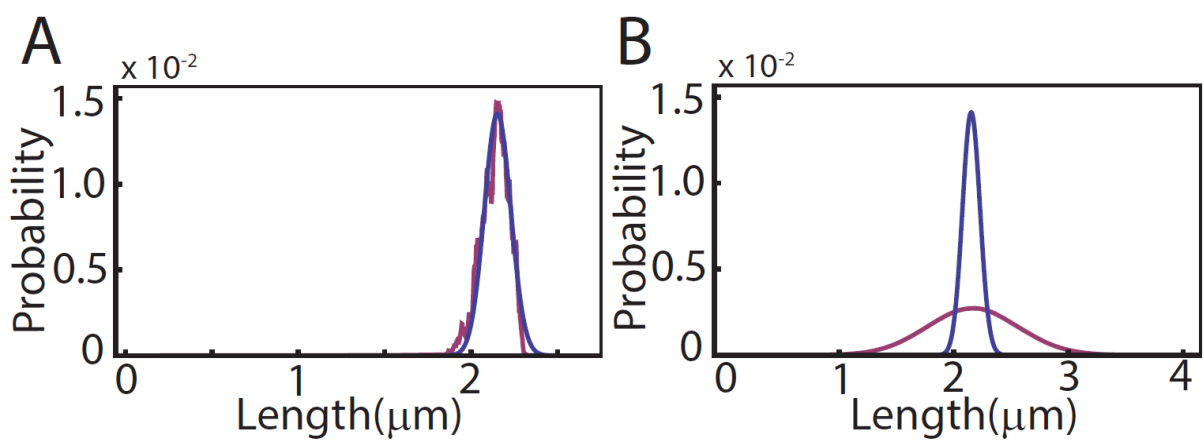

**Figure S1: Comparison of analytic and numerical distributions.** (A) In the fast switching regime the distribution obtained by detailed balance (blue) matches the cable length distribution obtained from the simulation (red). The parameters of the antenna model used to produce both distributions were $r = 0.2$, $d = 0.001$, $w = 10$, $k_{off} = 40$ all in units of s$^{-1}$. (B) When the rates of switching between the on (Smy1 not bound to formin) and off (Smy1 bound to formin) state are slow compared to the rates of polymerization and depolymerisation the analytic and numerical distribution differ. The means of the two distributions are the same while the correct distribution obtained numerically has a larger variance. The parameters used were $r = 200, d = 100, w = .005, k_{off} = 4$ all in units of s$^{-1}$.